# Thermodynamics of Ising Antiferromagnets with Phantom Cross-link Network on Husimi Lattice

Ran Huang1,2
1. Academy for Engineering and Applied Technology, Fudan University, Shanghai 200433, China
2. Department of Materials Technology and Engineering, Research Institute of Zhejiang University-Taizhou, Taizhou, Zhejiang 318000, China
Correspondence to huangran@fudan.edu.cn

A B S T R A C T

A second order cross-link network is set onto the classical Husimi lattice, to investigate the role of a "phantom" non-neighboring interactions of mid- and long-range in Bethe-like lattices for the first time. Since antiferromagnetic Ising model on Husimi lattice has been exactly solved and successfully presented the melting and glass transition, the Phantom Cross-link Network (PCN) is introduced here to understand the relationship between glassy defect and long-range interactions in small molecule systems, and the concept is inspired from the classical rubber network theory (Flory, 1985). One random site out of four on the recursive unites with certain distance $I$ (the net size) is selected to be linked onto the PCN. The solutions are still in the fashion of normal antiferromagnetic Ising model, with expected frustrations along with the net size. Beside the regular Curie transition, several interesting thermodynamics are observed in this toy model, and as the main found, PCN clearly introduces glassy portion into the system, identified by the supercooling behavior with lower $T_C$, the metastable entropy curve and the Kauzmann paradox.

**Introduction**

The Bethe-like recursive lattices have been developed to be powerful tools in the past decades [1,2], to approach the thermodynamic of various systems, such as classic Ising [3], elastic Ising [4], Potts [5], spin glass [2], metallic glass [6], biomacromolecules [7,8], etc. The fractal feature of such type of lattices has well-defined countable interactions and can avoid the complicity of many-body problem, which is usually dealt with mean field approximation, thereafter has been employed as an exactly calculable model [9]. Nevertheless, the claim that the particles in recursive lattice have the same circumstance due to the identical coordination number to the regular lattice in various dimensions remains controversial [3], one fundamental concern is that the single-node connectivity between any non-neighboring distance makes the graph with a strict hierarchy structure essentially a 1D system [10], and all the higher dimensions is quasi-approximated in local unit.

We applied a secondary "phantom" network of mid- and long range onto a classical antiferromagnetic Ising model on Husimi lattice, to investigate the recursive lattice with random non-neighboring interaction. As a toy model, the application of this Phantom Cross-link Network (PCN) is to overcome the 1D connectivity limitation of Bethe-like lattices, and extend the models to be more versatile. The term quoted "phantom network" in the rubber network theory [11], which originally means the polymer chains in rubber is not material and may pass through one another freely without volume exclusion [12]. Similar idea according in this work, the PCN is added free of constraints by neighboring spins, the junctions of the "phantom network" thus affected only by their connections to the network instead of the immediate surroundings. The methodology developed in this preliminary work have the potential to model the random field network, the medium-range order in glass forming [13], quasicrystal, or frustration in crystal [14].

**Model and methods**

A regular square lattice with an imaginary second order network with the size of three square-units (define the net size $I = 3$) is shown in Fig. 1a, one random site out of four on a linked unit (labeled as **P**) with certain distance $I$ is selected to be linked with an interaction $J_{pha}$ (marked as dash line). Analogically, Fig. 1b shows a partial structure of Husimi square lattice with the same idea of PCN, which is however an infinite fractal graph with an imaginary origin $O$. Notice here we select a random site out of the four on the linked unit to join the PCN, because the long-range interaction acting on a fixed site of periodic unit is trivial, which equals to a constant periodic outfield $H$, and subsequently, the sign of $J_{pha}$ has no effect (i.e. the PCN is randomly ferro- or antiferromagnetic). In this way, the Hamiltonian of the system is $\mathcal{H} = \sum_\alpha e_\alpha + \sum_p e_p = -J \sum_{ij} S_i S_j - J_{pha} \sum_{i,i+I} S_i S_{i+I}$, where $S_i = \pm 1$ indicate up and down spins, $J$ is set to be $-1$ as the reference exchange in this paper.

The calculation employs the "partial partition function" (PPF) technique and cavity method reported in previous works [2,3,15]. The detailed calculation code is provided in the supplementary materials [16]. The calculation starts with two initial guesses and iteratively approaches to a set of two steady solutions $X_1$ and $X_2$ (so called cycling solution), that is, the probability of a pair of neighboring sites occupied by a +1 spin, with the ratio computation of partial partition functions on each layer. Note that in the previous studies, only fixed-point 2-cycle solution $X_1$ and $X_2$ are concerned after the iterative calculations, in this study, since the presentation of random field the fixed point is inaccessible, we need to track the iterative progress of solutions with fluctuation to observe the effect of PCN, and take the mean value of steady $X$s.

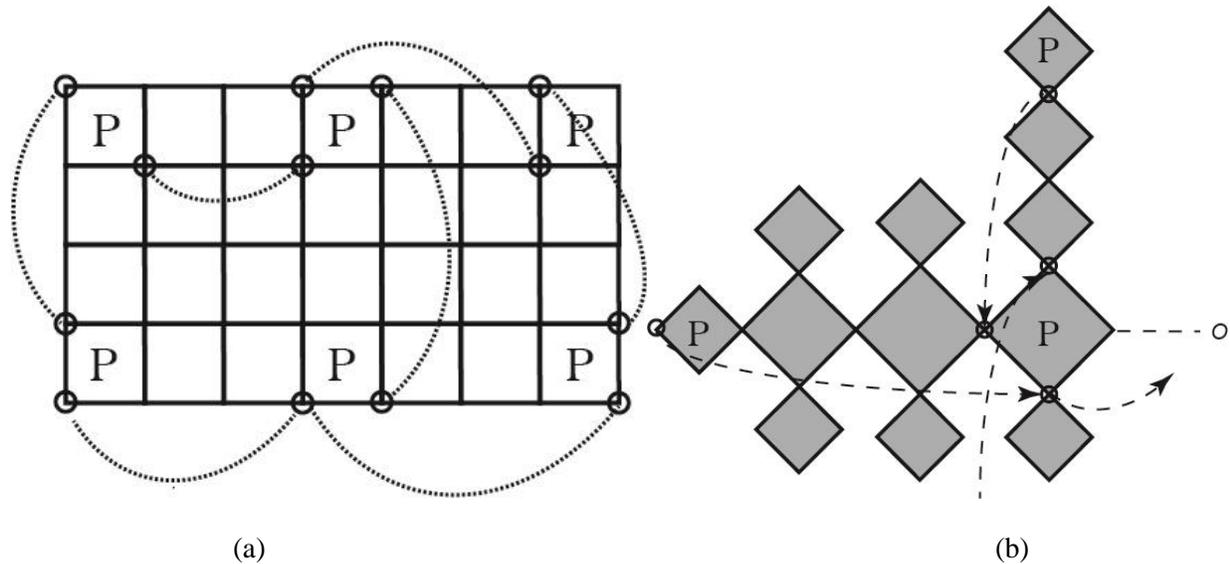

(a) (b)

Figure 1: Demonstration of a phantom cross-link network with $I = 3$ acting on (a) regular and (b) Husimi lattice.

## Results and discussion

The general solutions of $I = 3, 5, 10$, and $100$ with $J_{pha} = 1$ is presented in Fig. 2a. The spontaneous magnetization with Curie temperature $T_C$ is clearly observed with the solutions branching from 0.5 to 0 and 1 with the decrease of $T$ (indicating the probability of neighboring sites occupied by up or down spins). The difference with various $I$ is quite small, and the inset figure shows the details of the branch $X_2$ near below $T_C$, where tells the trend of $I$ from 3 to 100, that larger $I$ correlates to less frustration with higher stability as expected (the upper branch solution $X_2$ closer to 1). And nevertheless, the PCN does not shift the position of $T_C$. Referring to the calculation technique, the $X$s are iteratively approached to fixed-points along an imaginary direction pointing to the origin in the Husimi lattice, thus the $X$s affected by the PCN can be observed along with the calculation progress, while in deterministic cases we simply stop the calculation after the fixed point. Figure 2b shows the progressive solutions of $I = 100$ with $J_{pha} = 5$ along 1000 iteration steps after the fixed point at $T = 1.7$. We can see fluctuating values of $X$ at every 100 steps, and the derivation can be very large since $J_{pha}$ is large. It is easy to understand that with larger $I$ the frequency of irruptive $X$ is lower to affect the average solution and consequent thermal properties. In this way, we may take the difference between average progressive solutions $\tilde{X}$ and eigen solution $X$ (solution of the same system without PCN) to estimate the effectiveness of PCN. In contrast, Fig. 2c presents the progressive and average solutions of $I = 5$ PCN with $J_{pha} = 1$ at $T = 1.7$. It is observed that a phantom-linked site affecting about 5 units thereafter, thus the average $\tilde{X}$ distinguishes clearly from progressive $X$ in Fig. 2c, the overlapping effect is significant with $I \leqslant 5$. And for $I = 10$ the effect of PCN is not impressive as shown in Fig. 2d.

From Fig. 2b, we can see the upper effect of $J_{pha}$ with even outranged value is to reverse the $X$ to be $1-X$ on the other branch, therefore it is easy to understand the effect of $J_{pha}$ will be more violent as $X$ is smaller at lower $T$, and mild as $X$ approaching to 0.5 with $T \to T_C$, and we can estimate the upper limit of effects of $J_{pha}$ with various $I$ and $X$. Table 1 presents the theoretical upper limit of $\tilde{X}/X$ by the equation $\tilde{x} = \frac{(I-1)x+(1-x)}{Ix}$, which means for $I$ eigen $X$s, the upper effect of $J_{pha}$ reverses one $X$ to be $1-X$, then the ratio of average of $I-1$ $X$s and one $1-X$ to the eigen $X$ reflects the upper effect of PCN. If we roughly take 10% as an acceptable derivation, the up-right area in Table 1 is a reasonable range (marked grey) to have an effective PCN, i.e. the net size is not too small, and the temperature is not too low, otherwise the area with low temperature or very intensive PCN will destroy the regular order of the system and make the results trivial. Also, the analysis in table 1 agrees with two observations in Fig. 2: 1) the PCN does not affect the $T_C$ of spontaneous magnetization, 2) $I = 5 \sim 10$ is the meaningful range of network size of PCN.

To investigate the typical thermodynamics of PCN-affected HL, we take the conditions of $I = 5, J_{pha} = 1$ as an example to present the entropy behavior in Fig. 3: 3a shows five samples of the entropy of $I = 5$ PCN and the reference lattice without PCN, besides the $T_C$ are almost identical, the entropies of PCN-HL falls rapidly and lower than the reference at $T_K = 2.3$, gives a typical Kauzmann Paradox, and further falls negative at $T_{iK} = 1.7$, which is claimed to be the ideal glass transition that the following is unphysical and some transition must happen to avoid that [3]. This is an interesting implication, that at least in the present system, even amorphous fractures in the ordered state, i.e. glass embedded in crystal can undergo a significant ideal glass transition. And the thermals below $T_{iK}$ are trivial and not discussed further. On the other hand, the evidence of glassy region can also be traced with supercooled phenomenon, Fig. 3b emphasizes the entropies of PCN-HL near $T_C$. We applied a two-stage linear fitting on the entropy scatters due to the clear binding, which yields a lagged $T_C = 2.761327$ comparing to the referring $T_C = 2.770958$, followed by a state of higher entropy less stable than the pure crystal below the transition, that is, particles summoned by the PCN is out of the crystal-ordered domination and waiting for further cooling to drag the system down to another significant transition.

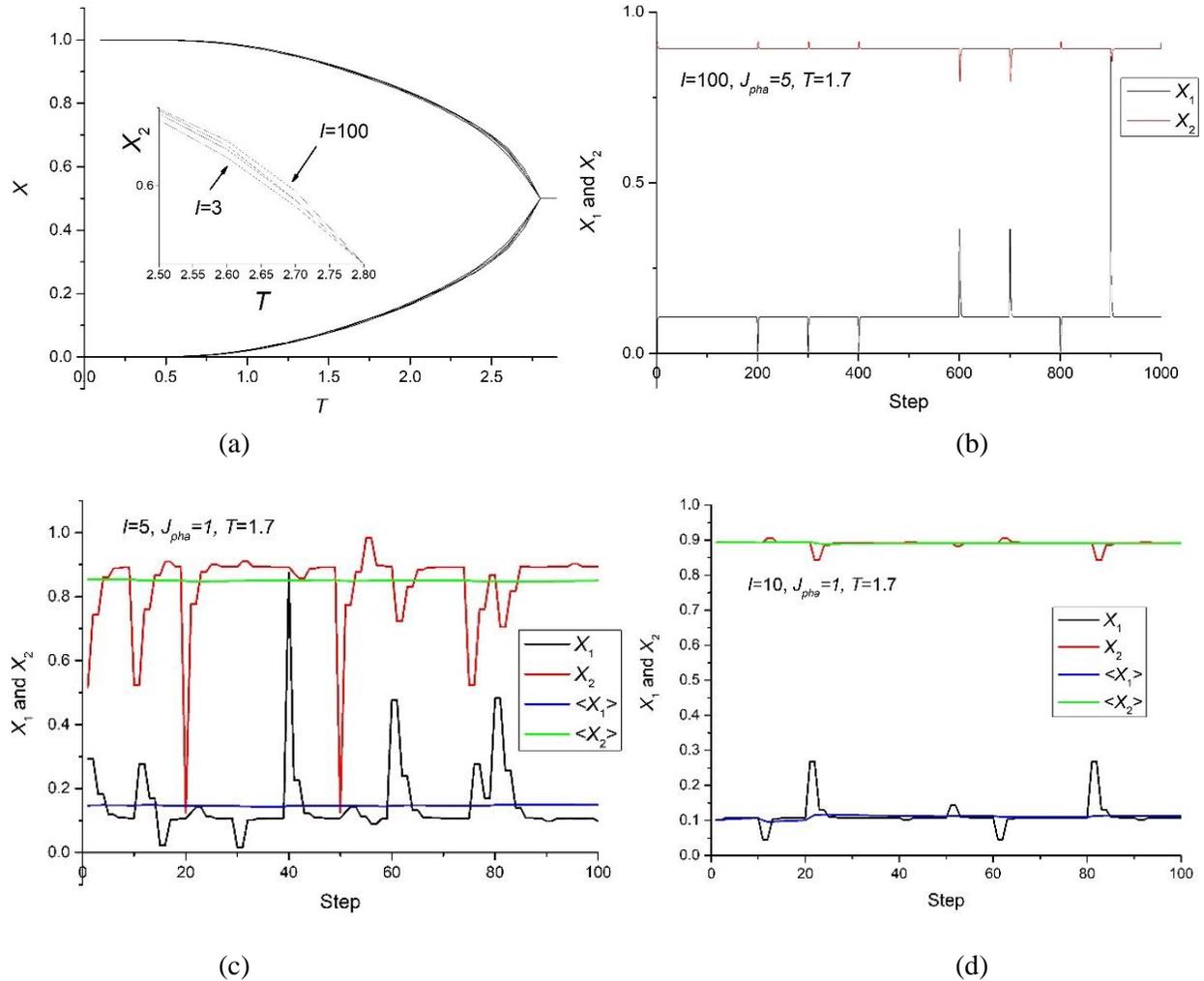

Figure 2: (a) The general solutions of $I = 3, 5, 10, 100$ with $J_{pha} = 1$. The inset is the details of the branch $X_2$ below $T_C$; (b) The progressive solutions of $I = 100$ PCN with $J_{pha} = 5$ along iteration steps; (c) The progressive and average solutions of $I = 5$ PCN with $J_{pha} = 1$; (d) The progressive and average solutions of $I = 10$ PCN with $J_{pha} = 1$.

**Conclusion**

The Phantom Cross-link Network is added onto classical Husimi lattice, to address the problem of single path connectivity of Bethe-like recursive lattices, which is somehow controversial on the practicality of recursive lattice to mimic the reality. This work establishes the modeling and calculation method, presents feasible results for the antiferromagnetic Ising model, and explores the effect of net size. It observes that the net size surprisingly does not affect the $T_C$, however leads to a Kauzmann Paradox as expected, since the PCN acts like glassy impurities in crystal. With the encouragement of these preliminary findings, this PCN@Husimi is believed to an interesting model, and we expect more insights to be unfolded, in either investigating the mystery of cross-dimensional Ising model, or serving as a powerful tool for general statistical systems.

Table 1: Theoretical estimation of upper limit effect of PCN: the value is derived from $\tilde{X}/X$; the gray area is a rough range of effective PCN.

| $I$ | $X = 0.01$ | $X = 0.05$ | $X = 0.1$ | $X = 0.2$ | $X = 0.3$ | $X = 0.4$ | $X = 0.49$ |
|---|---|---|---|---|---|---|---|
| 100 | 1.98 | 1.18 | 1.08 | 1.03 | 1.01 | 1.005 | 1.000408 |
| 50 | 2.96 | 1.36 | 1.16 | 1.06 | 1.07 | 1.01 | 1.000816 |
| 33 | 3.94 | 1.54 | 1.24 | 1.09 | 1.04 | 1.015 | 1.001224 |
| 25 | 4.92 | 1.72 | 1.32 | 1.12 | 1.05 | 1.02 | 1.001633 |
| 20 | 5.9 | 1.9 | 1.4 | 1.15 | 1.07 | 1.025 | 1.002041 |
| 10 | 10.8 | 2.8 | 1.8 | 1.3 | 1.13 | 1.05 | 1.004082 |
| 5 | 20.6 | 4.6 | 2.6 | 1.6 | 1.27 | 1.1 | 1.008163 |
| 4 | 25.5 | 5.5 | 3 | 1.75 | 1.33 | 1.125 | 1.010204 |
| 3 | 33.34 | 6.94 | 3.64 | 1.99 | 1.44 | 1.165 | 1.013469 |
| 2 | 50 | 10 | 5 | 2.5 | 1.67 | 1.25 | 1.020408 |

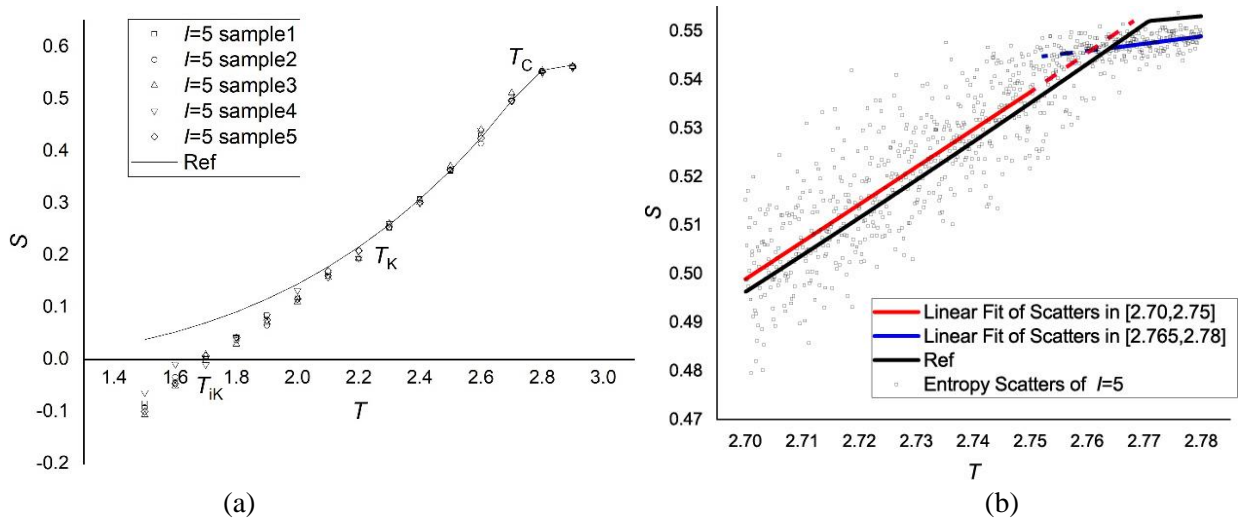

Figure 3: (a) The entropy of $I = 5$ PCN-HL and reference lattice; (b) the details of entropy behaviors around $T_C$.

**Acknowledgment**

This work is financially supported by the National Natural Science Foundation of China (11505110), the Medical Engineering Joint Fund of Fudan University (yg2021-005, yg2022-08), and the Taizhou Municipal Science and Technology Program (1801gy16, 2021zss04).